\documentclass{IEEEtran}  % Comment this line out if you need a4paper

\IEEEoverridecommandlockouts                              % This command is only needed if 
\makeatletter
\renewcommand{\@IEEEsectpunct}{.\ \,}% Modified from {:\ \,}
\makeatother
\usepackage{graphics} % for pdf, bitmapped graphics files
\usepackage{epsfig} % for postscript graphics files
\usepackage{mathptmx} % assumes new font selection scheme installed
\usepackage{times} % assumes new font selection scheme installed
\usepackage{amsmath} % assumes amsmath package installed
\usepackage{amssymb}  % assumes amsmath package installed
\usepackage[]{units}
\usepackage{gensymb}
\usepackage{placeins}
\usepackage{dblfloatfix}  
\usepackage{arydshln}

\title{\LARGE \bf
An Apparatus for the Simulation of Breathing Disorders\\
\Large{Physically Meaningful Generation of Surrogate Data}}

\author{Harry J. Davies, Ghena Hammour, Hongjian Xiao and Danilo P. Mandic\\
(harry.davies14, ghena.hammour17, d.mandic)@imperial.ac.uk% <-this % stops a space
}

\begin{document}

\maketitle
\thispagestyle{empty}
\pagestyle{empty}

%%%%%%%%%%%%%%%%%%%%%%%%%%%%%%%%%%%%%%%%%%%%%%%%%%%%%%%%%%%%%%%%%%%%%%%%%%%%%%%%
\begin{abstract}
The rapidly increasing prevalence of debilitating breathing disorders, such as chronic obstructive pulmonary disease (COPD), calls for a meaningful integration of artificial intelligence (AI) into healthcare. While this promises improved detection and monitoring of breathing disorders, AI techniques are almost invariably ``data hungry" which highlights the importance of generating physically meaningful surrogate data. Indeed, domain aware surrogates would enable both an improved understanding of respiratory waveform changes with different breathing disorders, and enhance the training of machine learning algorithms. To this end, we introduce an apparatus comprising of PVC tubes and 3D printed parts as a simple yet effective method of simulating both obstructive and restrictive respiratory waveforms in healthy subjects. Independent control over both inspiratory and expiratory resistances allows for the simulation of obstructive breathing disorders through the whole spectrum of FEV\textsubscript{1}/FVC spirometry ratios (used to classify COPD), ranging from healthy values to values seen in severe chronic obstructive pulmonary disease. Moreover, waveform characteristics of breathing disorders, such as a change in inspiratory duty cycle or peak flow are also observed in the waveforms resulting from use of the artificial breathing disorder simulation apparatus. Overall, the proposed apparatus provides us with a simple, effective and physically meaningful way to generate faithful surrogate breathing disorder waveforms, a prerequisite for the use of artificial intelligence in respiratory health.

%%%% intro

%breathing disorders - why is it important to simulate them? - the apparatus - results to verify it - conclusion
%%%

\end{abstract}

% ========================
% # I. Introduction      #
% ========================

\section{Introduction}

\IEEEPARstart{T}{he} prevalence of obstructive breathing disorders, such as chronic obstructive pulmonary disease (COPD) and asthma, is increasing rapidly \cite{Xie2020}, whilst other breathing disorders such as the restrictive pulmonary fibrosis (PF) continue to suffer from poor clinical outcomes and a lack of treatment options \cite{King2011}. Therefore, the understanding of breathing mechanics and resulting respiratory waveforms for different breathing disorders is paramount for the classification of breathing disorders, both in terms of screening and identifying their severity. To this end, we propose an apparatus for the artificial generation of obstructive breathing disorder waveforms through healthy subjects and mechanisms for reliably generating the whole spectrum of disease severities.

\subsection{Changes to breathing with obstruction and restriction}

Chronic obstructive pulmonary disease (COPD) is caused by an increased inflammatory response in the lungs which leads to obstructed airflow \cite{Viegi2007}. Chronic obstructive pulmonary disease encompasses both emphysema, defined by a breakdown in the elastic structure of the alveolar walls \cite{Thurlbeck1994} and bronchitis, defined by increased mucus secretion in the lungs \cite{Heard1979}. When we expire, the airways narrow due to reduced pressure, and thus if airway obstruction exists it is exaggerated during expiration. This explains why patients with COPD generally take longer to breath out than breathe in, and can generate higher inspiratory peak flows than expiratory peak flows.
The COPD can be diagnosed with a spirometry test, which measures the ratio of volume during forced expiration in one second (FEV\textsubscript{1}), against forced vital capacity (FVC). Practically those with COPD usually exhibit FEV\textsubscript{1} to FVC ratios of less than 0.7 \cite{Roman-Rodriguez2021}, but COPD is more specifically defined by different severities. According to the Global Initiative for Chronic Obstructive Lung Disease (GOLD) COPD can be split into 4 major categories based on FEV\textsubscript{1}. Mild COPD is defined as an FEV\textsubscript{1} of $\ge$ 80\% of a patients predicted FEV\textsubscript{1} based on height, age and sex. Moderate COPD is defined as an FEV\textsubscript{1} that is 50-79\% of it's predicted value, severe COPD is 30-49\% and finally very severe is defined as less than 30\% \cite{Patel2019}. The increased effects of obstruction during expiration also lead to a decreased inspiration time (T\textsubscript{I}) in comparison with the overall breathing time (T\textsubscript{TOT}) as it takes longer to breathe out. The ratio T\textsubscript{I}/T\textsubscript{TOT}, known as the inspiratory duty cycle, is lower in patients with COPD \cite{TOBIN1983286} \cite{Wilkens2010}.

This is in contrast to restrictive lung disease, an example of which is pulmonary fibrosis (scaring of the lungs). In this case, there is no obstruction of airways, but a restriction that applies equally to both inspiration and expiration. Whilst diagnosis of pulmonary fibrosis requires a multidisciplinary approach, such as the use of CT scans \cite{King2011}, spirometry tests will generally show healthy FEV\textsubscript{1}/FVC ratios, but with a lower peak flow for both inspiration and expiration as well as a greatly reduced vital lung capacity.

\begin{figure*}[h]
\centerline{\includegraphics[width=\textwidth]{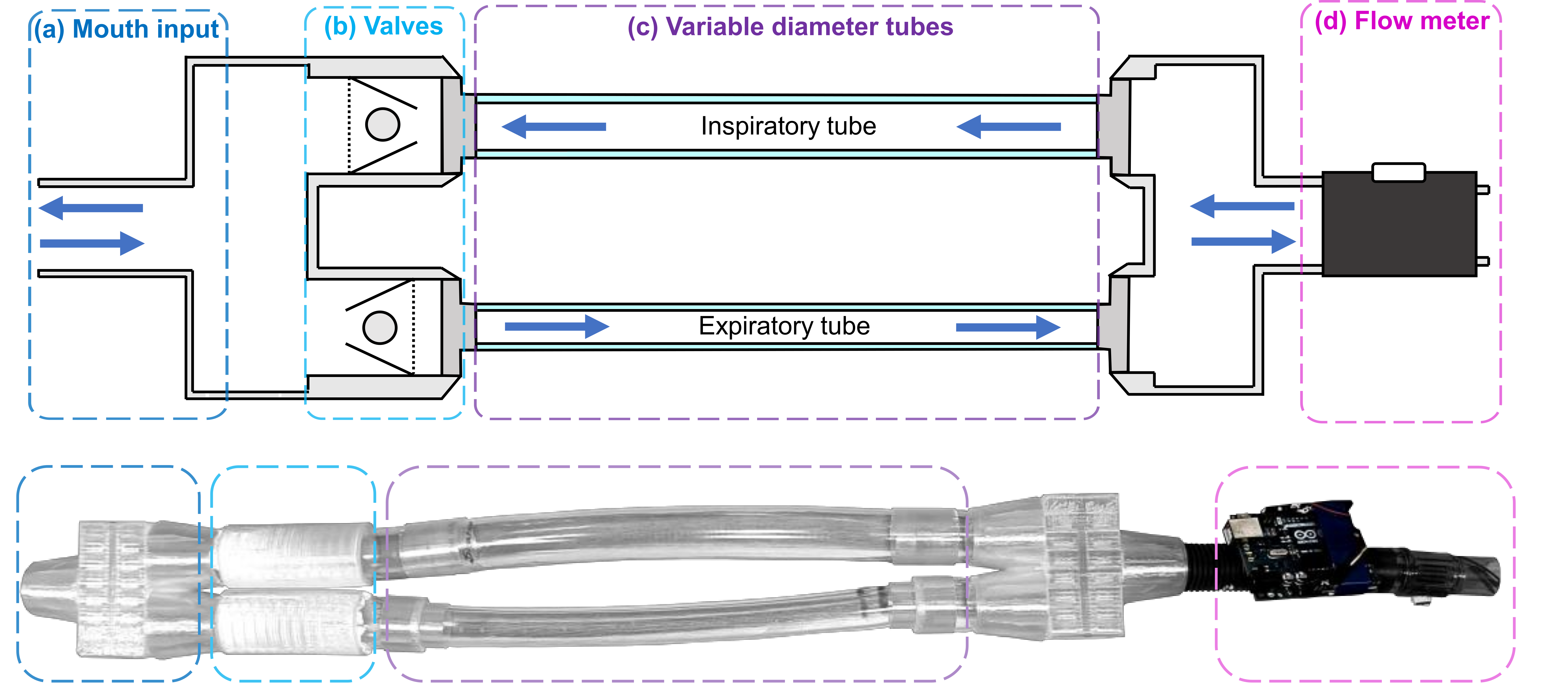}}
\caption{Block diagram (top) and physical realisation (bottom) of the proposed breathing disorder simulation apparatus. (a) The mouth input. (b) One-way valves in different directions for inspiration and expiration, comprised of a low density foam plug, a cone shaped funnel with a hole that is slightly smaller in diameter than the plug, and a fine mesh with allows air through but not the plug. (c) Tubes for both inspiration and expiration which can be easily swapped out for tubes of different diameter, allowing for independent control of resistances to inspiration and expiration. (d) A digital flow meter to record spirometry waveforms.}
\label{apparatus}
\end{figure*}

\begin{figure*}[h]
\centerline{\includegraphics[width=\textwidth]{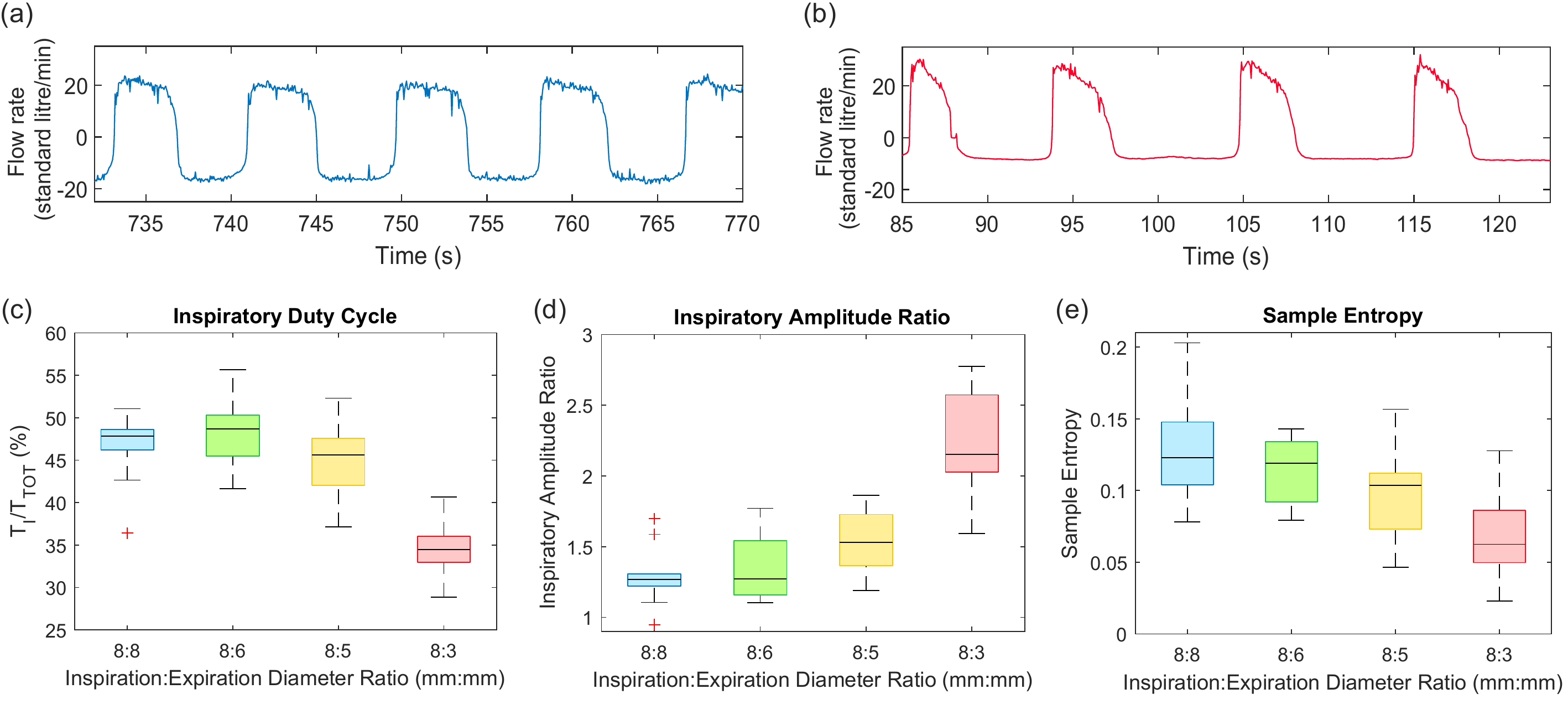}}
\caption{Tidal breathing results from using the apparatus with 4 different inspiration to expiration obstruction ratios across 10 subjects. (a) Exemplar spirometry waveform with an 8mm inspiratory tube and 8mm diameter expiratory tube giving a balanced obstruction ratio. Positive flow corresponds to inspiration and negative flow corresponds to expiration. (b) Exemplar spirometry waveform with an 8mm inspiratory tube and 3mm diameter expiratory tube giving an unbalanced obstruction ratio. (c) Boxplots of inspiratory duty cycle (\%), referring to the proportion of overall breathing duration spent in inspiration, across 10 subjects and 4 different inspiration:expiration tube diameter ratios. (d) Boxplots of inspiratory amplitude ratio, referring to peak inspiratory flow divided by peak expiratory flow, across 10 subjects and 4 different inspiration:expiration tube diameter ratios. (e) Boxplots of sample entropy (scale 1, tolerance = 0.2) across 10 subjects and 4 different inspiration:expiration tube diameter ratios.}
\label{spiro_results}
\end{figure*}

\begin{figure*}[h]
\centerline{\includegraphics[width=\textwidth]{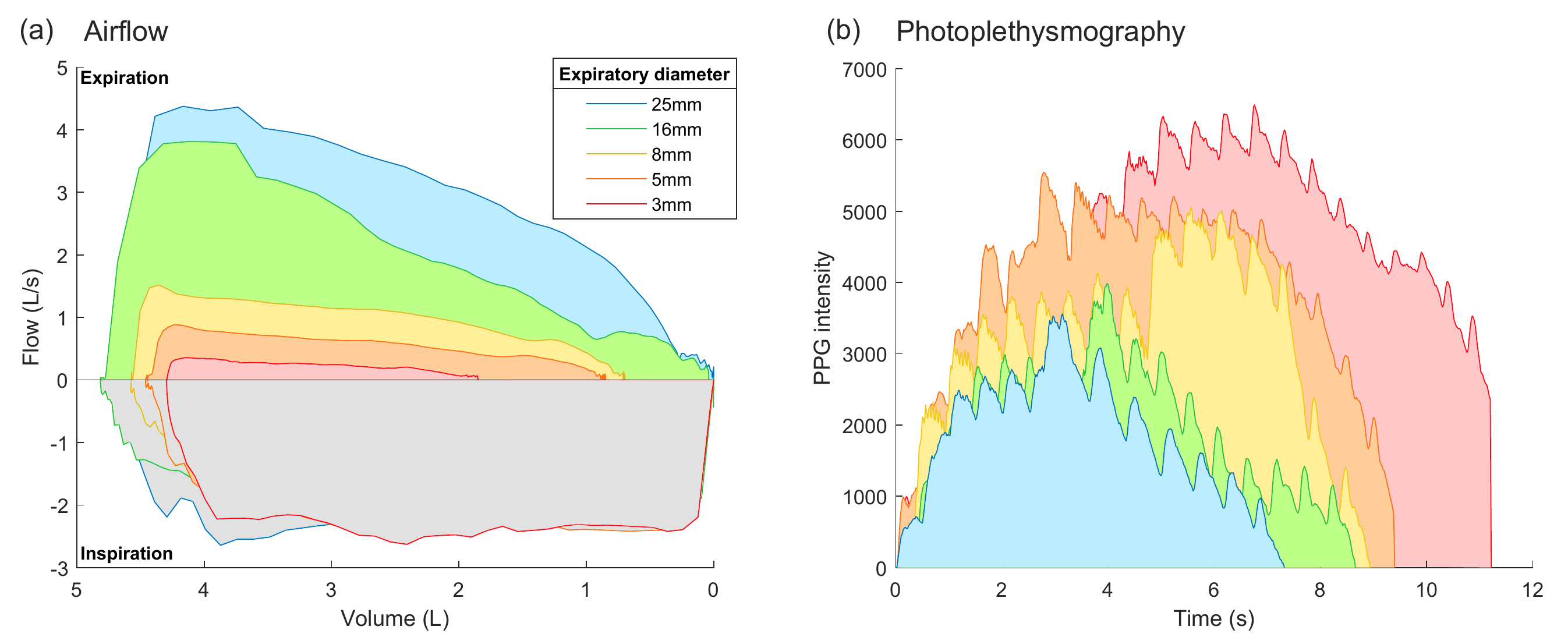}}
\caption{Example plots of maximally forced breathing for a single subject across 5 different expiratory tube diameters, ranging from 25mm to 3mm, with a fixed inspiratory tube diameter of 25mm providing low obstruction to inspiration. (a) Flow-volume loops for different expiratory tube diameters, showing a decreased flow for a given volume with a decrease in expiratory tube diameter. (b) Simultaneously recorded ear-photoplethysmography waveforms during maximally forced breathing with each tube diameter, showing an increase in both PPG intensity and duration with a decrease in tube diameter.}
\label{flow_ppg_loop}
\end{figure*}

\begin{figure}[h]
\centerline{\includegraphics[width=0.5\textwidth]{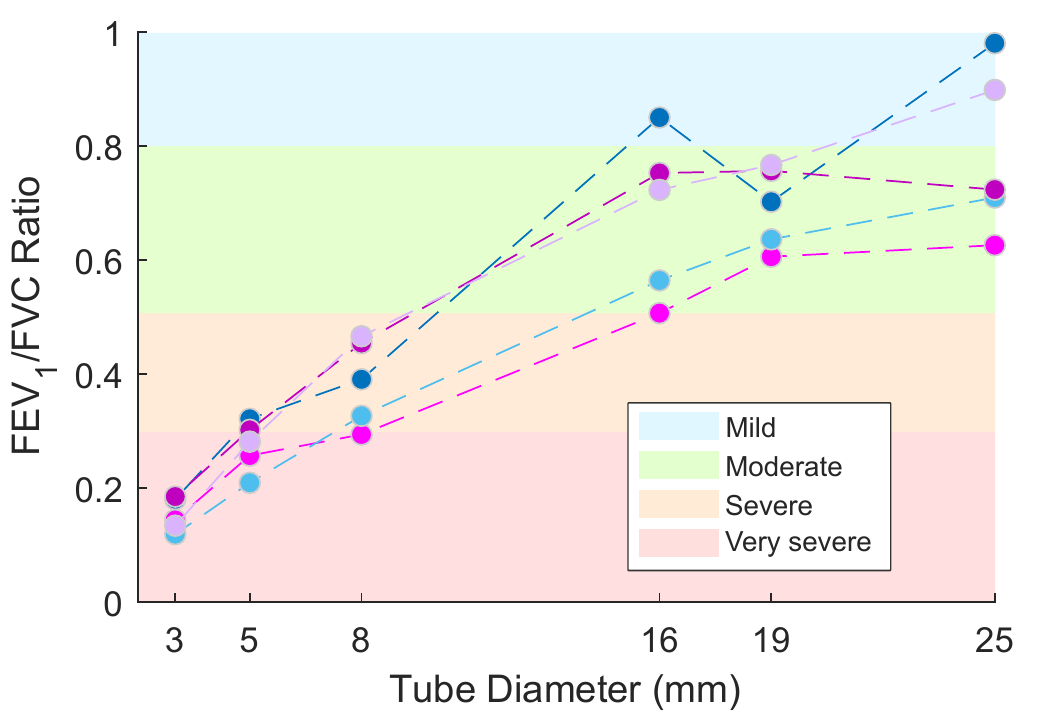}}
\caption{Calculated FEV\textsubscript{1}/FVC ratios across 6 expiratory tube diameters from 25mm to 3mm with an inspiratory tube diameter fixed at 25mm, plotted for 5 different subjects. Highlighted as shaded colours are the 4 different obstruction severities at the corresponding FEV\textsubscript{1}/FVC ratio, with blue indicating mild obstruction, green indicating moderate obstruction, orange indicating severe obstruction and red indicating very severe obstruction.}
\label{fev1_fvc_tubes}
\end{figure}

\subsection{Artificial changes to breathing resistance}

Resistance to breathing has been considered both to measure the strength and endurance of lungs in subjects, and also as a potential avenue to train lungs for increases in strength and endurance. A portable apparatus for collecting respiratory gas was designed in the early 1970s, comprising of tubes with 32mm diameter (giving negligible resistance to breathing) and a one-way valve so that gas could be stored when breathing out, but new air would be breathed in \cite{Daniels1971}. This apparatus was adapted in the mid to late 1970s by replacing the 32mm inspiratory tube with different smaller tube diameters (14mm, 11mm or 8mm), and breathing under different inspiratory resistances was examined in endurance athletes \cite{Dressendorfer1977}. A similar apparatus with four different inspiratory tube sizes was used to investigate the lung strength of a group of British coal miners over the age of 45 \cite{Love1977}. More recently, resistance has been applied to both inspiration and expiration through masks that have multiple inspiratory and expiratory valves, with the desire to train lungs for increased strength and endurance \cite{Kido2013}.

Different from the existing set-ups, the apparatus presented in this paper is capable of providing different resistances to both inspiration and expiration independently, with the aim of simulating the respiratory waveforms of different breathing disorders.

The so enabled simulation of breathing disorders through healthy subjects has the following benefits:
\begin{itemize}
  \item Ability to collect vast amounts data by expanding the subject pool to include healthy individuals;
  \item Full control over breathing resistances for both inspiration and expiration;
  \item Multiple obstructive breathing disorders of different severities can be investigated on the same healthy individual, thus keeping individual physiological differences constant;
  \item A controlled environment makes it easier to investigate how other physiological measures vary with resistance to breathing;
  \item A physically meaningful way to generate surrogate breathing disorder waveform data for both training and testing machine learning models.
\end{itemize}

%[J Daniels 1971] - 32mm hose sections - respiratory gas collection equipment - breathes into a one way valve which leads to gas collection bags.
 %[r dressendorfer 1977] adapted this equipment to change resistance to inspiration by altering the inpiratory tube diameter from 32mm, to either 14, 11 or 8mm diameter. shows large increases in pressure for a decrease in tube diameter when flow rate was kept constant.
 %[r love 1977] - varied inspiratory resistance in a similar manner with 4 different inspiratory tubes to investigate resistance to breathing in a group of coal miners over the age of 45.
 %[s kido 2013] - using a mask with two set inspiratory valves and two set expiratory valves to provide resistance to breathing during physical exertion.

%%

\section{Apparatus Design}
The apparatus consists of 3D printed parts and PVC tubes.  It has a single input tube which a subject breathes in and out of. This is connected to two one-way valves facing in opposite directions to switch the airflow path depending on inspiration and expiration. The valves consist of low density foam plug in a 3D printed cone shaped funnel with a hole slightly smaller than the diameter of the plug. Securing the ball in the funnel is a fine mesh in which air can pass through but the ball cannot. Depending on the orientation of the valve, either positive or negative airflow will seal the hole with the ball, thus preventing air from passing through. It is important that the plug is light so that it will move easily to the hole under low pressures. 

Connected to the inspiratory valve is an inspiratory tube which can be varied in diameter, as is the case with the expiratory valve and expiratory tube. The largest tube diameter is 25mm, which is considered as very low resistance to breathing. The smallest tube diameter used is 3mm, which provides very challenging resistance to breathing. To minimise the resistance of the whole apparatus, 3D printed parts also have an internal diameter of 25mm. Both the inspiratory and expiratory tubes are then connected to an output tube which leads into a SFM3200 digital flow meter by Sensiron (Stäfa, Switzerland) to record the breathing flow. The entire apparatus is shown in Fig.~\ref{apparatus}. The digital flow meter was connected to an Arduino Uno by Arduino (Somerville, MA, USA), which sampled flow values at a sampling frequency of 20Hz and displayed them on a computer monitor.

The apparatus was evaluated with tidal breathing in 10 subjects (5 male, 5 female) aged 18-30 years, across 4 different inspiration to expiration tube diameter ratios, and further evaluated with maximal forced breathing in 5 subjects (3 male, 2 female) across 6 different tube diameters for measurements of FEV\textsubscript{1}/FVC ratios.

Trial recordings were performed on 8 subjects (4 male, 4 female) aged 18-25 years, and included normal breathing under different resistances, as well as breathing in and out as hard as possible for both FEV\textsubscript{1}/FVC measurements and peak expiratory and inspiratory flow measurements. Photoplethysmography (PPG) was recorded from the ear simultaneously \cite{Davies2020} \cite{Davies2022}  during all recordings to gain insight into the effects of varying obstruction on thoracic pressure waveforms. Tidal PPG waveforms were then used to assess if a deep learning model that was trained exclusively on healthy and simulated disease data could be deployed to detect chronic obstructive pulmonary disease in real world PPG data.

The recordings were performed under the Imperial College London ethics committee approval JRCO 20IC6414, and all subjects gave full informed consent.

\section{Apparatus waveforms}

Small decreases in the expiratory tube diameter in relation to the inspiratory tube diameter resulted in changes to tidal breathing waveforms that are typical of patients with obstructive breathing disorders such as chronic obstructive pulmonary disease (COPD). Example spirometry waveforms in Fig.~\ref{spiro_results}(a,b) show the roughly symmetric breathing patterns when obstruction to inspiration and expiration is balanced (a) and the characteristic longer expiration time and reduced expiratory flow when obstruction to expiration is exaggerated with a tube diameter of only 3mm (b). Furthermore, these results are consistent across all 10 subjects, with Fig.~\ref{spiro_results}(c) showing a decrease in inspiratory duty cycle (percentage of overall breathing time spent inspiring) as obstruction to expiration is increased and Fig.~\ref{spiro_results}(d) showing an increase in the inspiratory amplitude compared with expiratory amplitude with increased obstruction. The median duty cycle of 34.4\% for the 8mm:3mm inspiration to expiration diameter ratio echoes the duty cycle of COPD patients, which were found to be around 35\% at rest \cite{TOBIN1983286}. Results also display a gradual decrease in sample entropy with increased obstruction to expiration, shown in Fig.~\ref{spiro_results}(e). Sample entropy is a measure of the complexity of a signal, and thus it is natural that sample entropy would decrease with increased obstruction, as obstruction decreases the degrees of freedom for breathing and in turn makes breathing patterns more predictable. Similar reductions in sample entropy have been shown in the breathing patterns of patients with COPD, with sample entropy decreasing as COPD severity increases \cite{Dames2014}.

The broad range of obstruction achievable by the apparatus is exemplified by the volume flow loops in Fig.~\ref{flow_ppg_loop}(a), which show decreased flow for a given volume with decreased tube diameter. This is specific to expiration due to the inspiratory tube diameter being kept constant whilst the expiratory tube diameter was varied, resulting in substantial changes to the expiration side of the volume flow loop with minimal changes to the inspiration side of the volume flow loop. It should be noted that the volume flow loops in Fig.~\ref{flow_ppg_loop}(a) also illuminate two important limitations of the apparatus. Firstly, whilst flow values are expected to decrease, overall recorded expiratory volumes should not decrease with tube diameter, given that this would not effect lung volume. The apparent reduction in recorded volume shown in the volume flow loops is due to leakage of the system at higher pressures, and could be rectified straight-forwardly with more robust materials and valves and joiners printed to more precise specifications. The second limitation is that the tube apparatus impedes breathing with a constant level of obstruction for a given tube diameter, whereas in reality as we expire the airways continue to narrow in proportion to lung volume decreasing. Obstruction in patients with COPD therefore increases further with continued expiration, resulting in a concave inflection in real-world volume flow loops that is not captured by this apparatus. 

Larger photoplethysmography (PPG) intensities are generated over a longer period of time with a decrease in expiratory tube diameter, as shown in Fig.~\ref{flow_ppg_loop}(b). Thoracic pressure increases as we expire to push air out of the lungs and this in turn decreases venous return to the heart and leads to the filling of peripheral venous beds at the site of the PPG probe. Increased PPG intensity thus reflects increased thoracic pressure and, as expected, thoracic pressure over time increases in proportion to increased obstruction simulated by smaller tube diameters, as increased pressure is required to force air through a smaller tube. Through measuring PPG, it is clear that the apparatus can simulate different thoracic pressure profiles internally based on changes in external obstruction. 

The apparatus was able to achieve a wide range of FEV\textsubscript{1}/FVC ratios across all subjects, with an example of the varied ratios in 5 subjects across 6 different expiratory tube diameters shown in Fig.~\ref{fev1_fvc_tubes}. The maximum FEV\textsubscript{1}/FVC achieved was 0.98 with the 25mm diameter tube, and the minimum achieved was 0.12 with the 3mm diameter tube. Importantly, artificially induced FEV\textsubscript{1}/FVC ratios were able to cover the full range of obstruction ratios across mild, moderate, severe and very severe. This indicates the promise of the tube based apparatus for simulating a full range of disease severities in each individual, and in turn vastly expanding the quantity of obstructive breathing disorder data available. 

\section{Surrogate COPD data for deep learning}

To test the abilities of the proposed apparatus for the generation of surrogate data, we recorded wearable in-ear photoplethysmography during tidal breathing with the apparatus and labelled it as chronic obstructive pulmonary disease for the purposes of training a convolutional neural network (CNN). After training on healthy PPG respiratory waveforms and artificially obstructed PPG respiratory waveforms that were labelled as COPD, the model was tested on a combination of unseen healthy data and real world COPD data. Detecting chronic obstructive pulmonary disease during tidal breathing from wearable PPG waveforms is difficult, considering that the transfer function from airflow in the lungs to PPG is a low pass filter and this vastly reduces the observable difference in features such as inspiratory duty cycle and skewness \cite{Davies2022}. Nevertheless, these timing and amplitude differences were present in both data generated by our and real-world data recorded from COPD patients. Respiratory waveforms were extracted from the photoplethysmography using multivariate empirical mode decomposition \cite{Rehman2010} with the three major PPG respiratory modes as inputs \cite{Davies2022}. To better facilitate the learning of the structural differences between healthy and diseased PPG, and not differences that occur based on respiratory rate or total amplitude changes resulting from individual sensor positioning, each segment of PPG respiratory cycles was standardised by dividing by the maximum absolute value, and all respiratory cycles were scaled to be 250 samples in length. We find that by training a deep learning model exclusively on PPG data recorded via the apparatus, that is labelled as COPD, the model was able to correctly classify unseen real world COPD data with high accuracy. These results will be presented in full in an upcoming publication.

\section{Conclusion}
We have demonstrated a simple yet effective method of simulating obstructive respiratory waveforms in healthy subjects with the use of a novel tube-based apparatus. Independent control over both in inspiratory and expiratory resistances allows for the simulation of respiratory waveforms corresponding to obstructive breathing disorders with a wide range of FEV\textsubscript{1}/FVC ratios, from healthy values through to values seen in very severe chronic obstructive pulmonary disease. Notably, this enables the investigation of obstructive breathing disorders at a range of severities in the same individual, allowing the waveform differences due to different tube resistances to be isolated whilst individual physiological differences are kept constant. Importantly, the proposed apparatus provides us with a physically meaningful way to generate surrogate breathing disorder waveforms, a prerequisite for testing and training machine learning models for the classification of breathing disorders.

% ==================
% # Acknowledgment #
% ==================
% use section* for acknowledgment
\section*{Acknowledgment}
This work was supported by the Racing Foundation grant 285/2018, MURI/EPSRC grant EP/P008461, and the Dementia Research Institute at Imperial College London.
%For the Summary paper submission only, no %acknowledgements are allowed. 

\FloatBarrier
% ==============
% # REFERENCES #
% ==============
%\section*{References}
\bibliographystyle{IEEEtran}
\bibliography{IEEEabrv,artificial}

% Generated by IEEEtran.bst, version: 1.14 (2015/08/26)
\begin{thebibliography}{10}
\providecommand{\url}[1]{#1}
\csname url@samestyle\endcsname
\providecommand{\newblock}{\relax}
\providecommand{\bibinfo}[2]{#2}
\providecommand{\BIBentrySTDinterwordspacing}{\spaceskip=0pt\relax}
\providecommand{\BIBentryALTinterwordstretchfactor}{4}
\providecommand{\BIBentryALTinterwordspacing}{\spaceskip=\fontdimen2\font plus
\BIBentryALTinterwordstretchfactor\fontdimen3\font minus
  \fontdimen4\font\relax}
\providecommand{\BIBforeignlanguage}[2]{{%
\expandafter\ifx\csname l@#1\endcsname\relax
\typeout{** WARNING: IEEEtran.bst: No hyphenation pattern has been}%
\typeout{** loaded for the language `#1'. Using the pattern for}%
\typeout{** the default language instead.}%
\else
\language=\csname l@#1\endcsname
\fi
#2}}
\providecommand{\BIBdecl}{\relax}
\BIBdecl

\bibitem{Xie2020}
M.~Xie, X.~Liu, X.~Cao, M.~Guo, and X.~Li, ``{Trends in prevalence and
  incidence of chronic respiratory diseases from 1990 to 2017},''
  \emph{Respiratory Research}, vol.~21, no.~1, p.~49, Feb 2020.

\bibitem{King2011}
T.~E. King, A.~Pardo, and M.~Selman, ``{Idiopathic pulmonary fibrosis},''
  \emph{The Lancet}, vol. 378, no. 9807, pp. 1949--1961, Dec 2011.

\bibitem{Viegi2007}
G.~Viegi, F.~Pistelli, D.~L. Sherrill, S.~Maio, S.~Baldacci, and L.~Carrozzi,
  ``{Definition, epidemiology and natural history of COPD},'' \emph{European
  Respiratory Society}, vol.~30, no.~5, pp. 993--1013, Nov 2007.

\bibitem{Thurlbeck1994}
W.~M. Thurlbeck and N.~L. M{\"{u}}ller, ``{Emphysema: Definition, imaging, and
  quantification.}'' \emph{American Journal of Roentgenology}, vol. 163, no.~5,
  pp. 1017--1025, 1994.

\bibitem{Heard1979}
B.~E. Heard, V.~Khatchatourov, H.~Otto, N.~V. Putov, and L.~Sobin, ``{The
  morphology of emphysema, chronic bronchitis, and bronchiectasis: Definition,
  nomenclature, and classification.}'' \emph{Journal of Clinical Pathology},
  vol.~32, no.~9, p. 882, 1979.

\bibitem{Roman-Rodriguez2021}
M.~Roman-Rodriguez and A.~Kaplan, ``{GOLD 2021 strategy report: Implications
  for asthma–COPD overlap},'' \emph{International Journal of Chronic
  Obstructive Pulmonary Disease}, vol.~16, p. 1709, 2021.

\bibitem{Patel2019}
A.~R. Patel, A.~R. Patel, S.~Singh, S.~Singh, and I.~Khawaja, ``{Global
  initiative for chronic obstructive lung disease: The changes made},''
  \emph{Cureus}, vol.~11, no.~6, Jun 2019.

\bibitem{TOBIN1983286}
M.~J. Tobin, T.~S. Chadha, G.~Jenouri, S.~J. Birch, H.~B. Gazeroglu, and M.~A.
  Sackner, ``{Breathing patterns: 2. Diseased subjects},'' \emph{Chest},
  vol.~84, no.~3, pp. 286--294, 1983.

\bibitem{Wilkens2010}
H.~Wilkens, B.~Weingard, A.~{Lo Mauro}, E.~Schena, A.~Pedotti, G.~W. Sybrecht,
  and A.~Aliverti, ``{Breathing pattern and chest wall volumes during exercise
  in patients with cystic fibrosis, pulmonary fibrosis and COPD before and
  after lung transplantation},'' \emph{Thorax}, vol.~65, no.~9, pp. 808--814,
  2010.

\bibitem{Daniels1971}
J.~Daniels, ``{Portable respiratory gas collection equipment},'' \emph{Journal
  of Applied Physiology}, vol.~31, no.~1, pp. 164--167, Jul 1971.

\bibitem{Dressendorfer1977}
R.~H. Dressendorfer, C.~E. Wade, and E.~M. Bernauer, ``{Combined effects of
  breathing resistance and hyperoxia on aerobic work tolerance},''
  \emph{Journal of Applied Physiology}, vol.~42, no.~3, pp. 444--448, 1977.

\bibitem{Love1977}
R.~G. Love, D.~C. Muir, K.~F. Sweetland, R.~A. Bentley, and O.~G. Griffin,
  ``{Acceptable levels for the breathing resistance of respiratory apparatus:
  results for men over the age of 45.}'' \emph{Occupational and Environmental
  Medicine}, vol.~34, no.~2, pp. 126--129, May 1977.

\bibitem{Kido2013}
S.~Kido, Y.~Nakajima, T.~Miyasaka, Y.~Maeda, T.~Tanaka, W.~Yu, H.~Maruoka, and
  K.~Takayanagi, ``{Effects of combined training with breathing resistance and
  sustained physical exertion to improve endurance capacity and respiratory
  muscle function in healthy young adults},'' \emph{Journal of Physical Therapy
  Science}, vol.~25, no.~5, pp. 605--610, May 2013.

\bibitem{Davies2020}
H.~J. Davies, I.~Williams, N.~S. Peters, and D.~P. Mandic, ``{In-ear
  SpO\textsubscript{2}: A tool for wearable, unobtrusive monitoring of core
  blood oxygen saturation},'' \emph{Sensors}, vol.~20, no.~17, 2020.

\bibitem{Davies2022}
H.~J. Davies, P.~Bachtiger, I.~Williams, P.~L. Molyneaux, N.~S. Peters, and
  D.~Mandic, ``{Wearable in-ear PPG: Detailed respiratory variations enable
  classification of COPD},'' \emph{IEEE Transactions on Biomedical
  Engineering}, vol.~69, no.~7, pp. 2390--2400, 2022.

\bibitem{Dames2014}
K.~K. Dames, A.~J. Lopes, and P.~L. {De Melo}, ``{Airflow pattern complexity
  during resting breathing in patients with COPD: Effect of airway
  obstruction},'' \emph{Respiratory Physiology and Neurobiology}, vol. 192,
  no.~1, pp. 39--47, Feb 2014.

\bibitem{Rehman2010}
N.~Rehman and D.~P. Mandic, ``{Multivariate empirical mode decomposition},''
  \emph{Proceedings of the Royal Society A: Mathematical, Physical and
  Engineering Sciences}, vol. 466, no. 2117, pp. 1291--1302, May 2010.

\end{thebibliography}

\end{document}